\documentclass[aps,prl,preprint,groupedaddress,showpacs]{revtex4-1}
\usepackage{amsmath,bm}
\usepackage{amssymb}

\usepackage{graphics}
\usepackage{epstopdf}
\usepackage{epsf}

\def\ve{\varepsilon}

\def\p{\partial}

\def\sun{\odot}
\usepackage{graphicx}
\usepackage{graphics}
\usepackage{epstopdf}
\usepackage{epsf}
\newcommand\aap{A\&A}                
\newcommand\aj{AJ}                   

\newcommand\apjl{ApJ}                
\newcommand\apss{Ap\&SS}             
\newcommand\cjaa{Chinese J.~Astron. Astrophys.} 
\newcommand\mnras{MNRAS}             
\begin{document}
\title{Ratio of kinetic  luminosity  of the jet to bolometric luminosity of the disk  at the   ``cold'' accretion onto a supermassive black hole}

\author{S.V.Bogovalov}

\affiliation{National Research Nuclear University (MEPHI), Kashirskoje shosse, 31, Moscow, Russia}

\date{25.06.2018}



\begin{abstract}
Disk accretion onto black holes is accompanied by collimated outflows (jets). In active galactic nuclei (AGN), the  kinetic energy flux of the jet may exceed the  bolometric luminosity of the disk a few orders of magnitude. This phenomena can be explained in frameworks of so called ``cold'' disk accretion when the only source of energy of AGN  is the energy released at the accretion.  The  radiation from the disk is suppressed 
because the wind from the disk carries out almost all  the angular momentum and the gravitational energy of the accreted material.  In this paper, we 
calculate  an unavoidable radiation from the ``cold'' disk and the ratio of the kinetic energy luminosity of the outflow to the bolometric luminosity of  the accretion disk around a super massive black hole in  framework of the paradigm of the optically thick $\alpha$-disk of Shakura \& Sunyaev. The ratio of the luminosities is defined predominantly by the  ratio of the magnetic field pressure inside the disk to the magnetic field pressure at the base of the wind.  The obtained equations applied to the jet of  M87 demonstrate good agreement with observations.  In the case of Sgr A*,  these  equations allow us to predict the kinetic energy flux from the disk around Galactic SMBH.  
\end{abstract}

\maketitle
\section{Introduction}
In the standard  theory of disk accretion  onto a black hole  \cite{1973A&A....24..337S} 
is assumed that  the gravitational energy of the accreted material is entirely carried out by photons radiated from the surface of the optically thick disk.  Measurements of the bolometric luminosity of the disk gives total energy released at the accretion. 
However, the disk accretion can be accompanied by formation of collimated outflows (jets). 
Moreover, the observations of recent years show that the power of kinetic energy  of the jets in some  AGN   exceeds the bolometric luminosity of the disk. 

The famous galaxy M87 is a striking example  of an AGN with a very large  kinetic-to-bolometric luminosity. According to \cite{1991AJ....101.1632B},  the bolometric luminosity of the disk does not exceed $10^{42}~\rm erg/s$ while the kinetic luminosity of the jets is of the order of 
$10^{44} ~\rm erg/s$ \cite{1996ApJ...467..597B,1996MNRAS.283L.111R}.  

Starting with the paper  \cite{1991Natur.349..138R}, the  energetics and lifetimes of extended double radio sources have been  used for calculation of the  jet power  in radio galaxies
and quasars.   The ratio of kinetic-to-bolometric luminosity can be estimated from radio and X-ray data. \cite{2003MNRAS.343L..59H,2003MNRAS.345.1057M,2004A&A...414..895F,2006MNRAS.369.1451K,2007MNRAS.381..589M,2011MNRAS.415.2910D,2015MNRAS.450.2317S}   argued that the radio and X-ray luminosities are likely to be related
to  the kinetic and bolometric luminosities, respectively. 
It is established that   in large fraction of AGNs  the jet kinetic luminosity exceeds the 
bolometric luminosity   \cite{2008MNRAS.383..277K,2008ChJAA...8...39M,2012A&A...544A..56L,2016MNRAS.458L..24D,2016arXiv160601399D,2011MNRAS.411.1909F,2011ApJ...728L..17P}.

The jet power in 191 quasars detected by the Fermi
Large Area Telescope (LAT) in gamma rays,   
calculated  within the  frameworks of an one-zone  model  show that in this sample of blazars  $L_{kin}$ systematically exceeds the bolometric luminosity \cite{2014Natur.515..376G}.

An indirect evidence  of high kinetic luminosity of an outflow exceeding the bolometric luminosity  is provided by observations of the Galactic Center   in  TeV gamma-rays \cite{2016Natur.531..476H}. 
To explain the observed diffuse flux of the VHE gamma-rays from the Galactic Center region, 
the production  rate of  protons accelerated up to  1~PeV,   should be  
$\sim 10^{38}~ \rm erg/s$.  Assuming that the accelerator of protons is powered by the kinetic energy of the outflow (a wind or jets)  from the super massive Black Hole in the 
the Galactic Center (Sgr A*),  even in the case of 100  \% conversion of the bulk kinetic energy to non thermal particles,   the kinetic luminosity of the outflow would two orders of magnitude exceed the bolometric luminosity of Sgr A* which is estimated close to 
 $10^{36}~\rm erg/s$ \cite{2010RvMP...82.3121G}.

Observations of the very powerful and bright in gamma-rays AGN  3C 454.3 
give even more interesting information.  During the outbursts of this object, 
its  apparent luminosity in GeV gamma-rays  could exceed  $10^{50}~ \rm erg/s$ 
\cite{2010ApJ...718..455S,2010ApJ...721.1383A,2011ApJ...736L..38V,2011ApJ...733L..26A}.
The mass of the black hole in this AGN is estimated in the region $(0.5 - 4)\cdot 10^9~\rm M\sun$.
Thus  the  Eddington luminosity is in the range  of $(0.6 - 5) \cdot 10^{47} ~\rm erg/s$.
Because of the Doppler boosting effect, the intrinsic gamma-ray luminosity of this source is 
much smaller, by several orders of magnitude, than 
the apparent luminosity. Yet, the estimates  of the jet kinetic luminosity in an any 
realistic scenario  give a value exceeding the Eddington luminosity \cite{2013ApJ...774..113K}. 

In general, the estimates of the bolometric and kinetic luminosities are model dependent \cite{2017MNRAS.465.3506P}. Nevertheless, it is difficult to avoid a conclusion that  at least some AGN demonstrate extremely high kinetic luminosities of jets which are not only 
above the bolometric luminosity, but in some cases can  exceed the Eddington luminosity 
of the central super massive black hole (SMBH).   
In this case the question about the source of the energy of the jets becomes the central one for understanding of the  processes at the accretion onto SMBH \cite{2011ApJ...727...39M}. 

Presently,  the 
rotational energy of a black hole is considered as the most likely source of energy which is transformed in to the energy of jets due to 
the so-called mechanism of Blandford and Znajek  \cite{1977MNRAS.179..433B}.
In this scenario,  SMBH  provides an additional (to the accretion) source of energy 
which, in fact,  could be the dominant  source of energy of the system. According to numerical simulations \cite{2011MNRAS.418L..79T,2012MNRAS.421.1351B} this mechanism can provide the energy flux in the jet  $\approx 3\dot M c^2$ , where $\dot M$  is the accretion rate. But  the black hole has to rotate with maximal possible angular momentum. In this case the 
radiation luminosity of the optically thick disk can achieve $\eta \dot M c^2$ with $\eta =0.3$. Thus, the  maximal ratio of the  kinetic-to-bolometric luminosities is close to 10. This value can be increased  if the accretion occurs in radiatively inefficient regime \cite{1976ApJ...204..187S,1982Natur.295...17R,1995ApJ...452..710N}. In this regime the luminosity of the disk is suppressed because the energy released at the  accretion is predominantly advected in to the black hole. 
But the alternative model is also possible in which the only source of the energy of AGN is the energy released at the accretion .

It was firstly pointed out in \cite{1992ApJ...394..117P} that the magnetized wind from the disk can carry out a significant fraction of the angular momentum of the accreted matter rather than the viscous stresses. This splendid idea  has been explored in a range of  papers by  Grenoble  group \cite{1993A&A...276..625F,1995A&A...295..807F,1997A&A...319..340F} which called this type of flow around black holes as Magnetized Accretion-Ejection structures (MAES). A range of other authors last years explored similar approach \cite{2013ApJ...765..149C,2014ApJ...786....6L,2014ApJ...788...71L} in different physical context.  In the last works of the Grenoble group 
\cite{2018arXiv180304335M,2018arXiv180512407M} the radiation from the Jet Emitting Disks (JED) is discussed. But the discussion is limited by the case when the jet carries out a moderate fraction (20\%-80\%) of the energy released at the accretion. 

Starting with our first work \cite{2010IJMPD..19..339B} devoted to this problem we stressed that the idea 
of \cite{1992ApJ...394..117P} can be applied to the solution of the problem of high ratio of the kinetic luminosity of the jets to the bolometric luminosity of the disks. 
Actually  \cite{1995A&A...295..807F,2006A&A...447..813F}  specially stressed much before us that at the conventional conditions the jet can carry out almost all the angular momentum and energy from the accretion disk. The radiation of the disk can carry out only a minor fraction of the accretion energy. Therefore, the ratio of the kinetic luminosity of the jets on a few orders of magnitude can exceed the bolometric luminosity of the disks in this regime of accretion.   
We called this type of accretion as 
"dissipationless`` or ''cold`` accretion because the dissipative processes like turbulent viscosity play no role in the dynamics of the accretion. Two questions arise in this regard:
\begin{itemize}
 \item Do solutions describing the regime of ''cold`` accretion exist?
 \item If yes  then what is the  expected ratio of the kinetic-to-bolometric luminosity? Is it really possible to have very high ratio of the kinetic-to bolometric luminosity exceeding 100? 
\end{itemize}

Presently,  we have a positive answer to  the first question. The selfconsistent solution of the problem of accretion and outflow has been obtained in \cite{1995A&A...295..807F} in the self similar approximation. This allowed to the authors to calculate the structure of the accretion disk assuming that the 
magnetic field is formed due to advection and diffusion \cite{1997A&A...319..340F,2000A&A...353.1115C}.

In the work by \cite{2010IJMPD..19..339B} a self similar solution of the problem of ''cold`` accretion has been obtained neglecting   the structure of the accretion disk. From one side this  simplified the model but from other side this approach  allows us to develop a method of numerical solution of the problem without self similarity assumptions and obtain more general numerical solutions 
\cite{2017AstL...43..595B,doi:10.1142/S0218271818440054}. All these works remain no doubts that the ''cold`` accretion due to magnetized wind  is 
really possible.

In this paper we try to  answer to  the  second question. ''Cold`` or Jet Emitting Disks \cite{2006A&A...447..813F}  can formally exist without any dissipation processes (viscosity and finite electric conductivity) and therefore formally can remain cold. But in reality 
the heating of the disk is unavoidable because the ''cold`` disk accretion is possible only in presence of  magnetic field at the base of the wind. Therefore, the magnetic field has to be present also inside the disk producing magnetic and turbulent viscosity. The objective of the work is to estimate the heating of the disk due to    
turbulence and determine the expected level of the ratio of the kinetic-to-bolometric luminosity of the ''cold``  disks.

\section{Physics  of  the ''cold`` accretion}

\begin{figure}
\centering
 \includegraphics[width=0.7\textwidth]{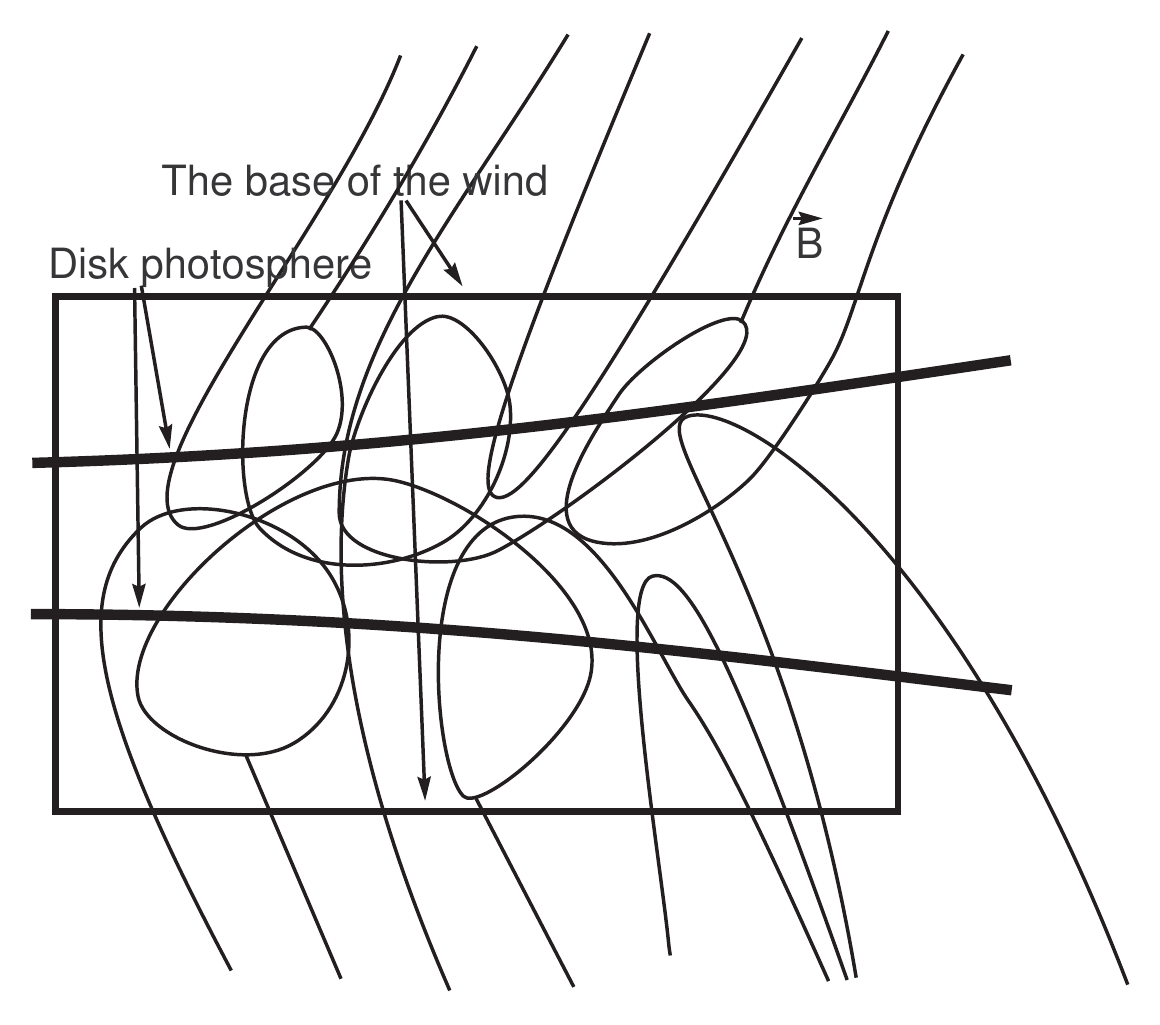}
 \caption{Schematic structure of the magnetic field inside the disk and in the wind. The wind is formed in the corona where all field lines are open. Inside the disk, the chaotic magnetic field is generated due to  turbulent motion of plasma in the disk.}
\label{fig0}
\end{figure}

In the work of the Grenoble group the magnetic field in the disk and outflow has formed under advection and diffusion of the external regular field \cite{1976Ap&SS..42..401B,2012ApJ...750..109B}. The magnetic field lines predominantly vertically crosses the plane of the disk.  But this is apparently strongly simplified case. Following to \cite{1973A&A....24..337S}, we assume that the magnetic field of the accretion disk is generated by the turbulent motion of plasma  due to a dynamo mechanism \cite{2001ApJ...548L...9M,2003ApJ...593..184L,2004MNRAS.348..111K,2014ApJ...782L..18B,2015ApJ...809..118B,2016MNRAS.457..857S,2017A&A...598A..87R}. The formation of regular vertical component of the magnetic field occurs due to emerging, reconnection  \cite{1979ApJ...229..318G} and  opening of magnetic loops by the disk differential rotation \cite{1998ApJ...500..703R}. The magnetic field structure inside the disk is shown schematically in fig.\ref{fig0}. The wind  is formed  quite high, in the corona where all the field lines are  open, and the magnetic pressure $B^2/8\pi$ exceeds the thermal pressure $p$.  The outflow of plasma from the disk occurs due to centrifugal \cite{1982MNRAS.199..883B} mechanism.

The equation of conservation of the angular momentum can be obtained directly from  MHD equations integrated  over the central volume shown in the form of rectangular box in fig.\ref{fig0} (for technical details see \cite{2010IJMPD..19..339B}).  This equation in framework of the 
$\alpha$-disks paradigm \cite{1973A&A....24..337S} has the following  form: 
\begin{equation}
 \dot M {\partial r V_k\over \partial r}+{\partial \over \partial r}4\pi r^2t_{r\varphi} h+r^2 <B_{\varphi} B_z>|_{wind}=0,
 \label{e0}
\end{equation}
where $t_{r\varphi}$ is the tangential stress, $V_k$ is the Keplerian velocity of rotation, $\dot M$ is the accretion rate,  $r$ is the cylindrical radius and $h$ is the semi-height of the disk.  $B_{\varphi}$ and  $ B_z$ are  
toroidal and z-components of the magnetic field at the base of the wind. The 
brackets $< >$ mean  averaging over time. The term ${\partial \over \partial r}4\pi r^2t_{r\varphi} h$ is responsible for the viscous loss of the angular momentum by the disk. The last term in the left hand side of the equation is responsible for the angular momentum loss due to the magnetized wind.  Symbol $|_{wind}$ means that the value is taken at the base of the wind at the surface of the disk.

The tangential stress in the disk $t_{r\varphi}$ is defined by the  turbulent motion  and by the magnetic field. According to  \cite{1973A&A....24..337S}
\begin{equation}
t_{r\varphi}=-\alpha\rho v_s^2.
\end{equation}
where $v_s$ is the sound velocity, $\rho$ is the matter density and $\alpha$ is the parameter connecting the pressure and tangential stress in the disk. 

As it was pointed out by \cite{1992ApJ...394..117P}, the momentum loss due to the wind  will dominate  the losses caused by the  viscous stresses provided  that 
\begin{equation}
 4\pi rt_{r\varphi} h \ll r^2 <B_{\varphi} B_z>|_{wind}.
 \label{iq1}
\end{equation}
In the opposite case we have the standard \cite{1973A&A....24..337S} version of the disk accretion.

The integration of the energy conservation equation over the control volume  gives 
\begin{equation}
{\p\over \p r} \dot M {V_k^2\over 2}-{\partial \over \partial r}4\pi\Omega r^2t_{r\varphi} h+4\pi r\rho V_z E_t|_{wind}+4\pi r Q=0,
\label{e01}
\end{equation}
where
\begin{equation}
 E_{tot}={v^2\over 2} -{B_z B_{\varphi}\over 4\pi \rho V_z}\Omega r  - {GM\over r}
\label{e02}
\end{equation}
is the total energy per particle in the wind, $Q$ is the energy radiated from unit square of one side of the disk surface.
We neglect by  the contribution from  the turbulent Ohmic heat production  and  advection of the energy in the optically thick disk.  

Taking into account the mass conservation
\begin{equation}
 {\partial \dot M\over\partial r}-4\pi r\rho v_z |_{wind}=0,
 \label{e4}
\end{equation}
a simple algebra with eqs. (\ref{e0}) and (\ref{e01}) gives
like in \cite{1973A&A....24..337S},  that
the heat production equals 
\begin{equation}
 Q=t_{r\varphi}hr{\p\Omega\over \p r}. 
 \label{e03}
\end{equation}

Below we consider the case when the inequality in eq.(\ref{iq1}) is fulfilled.   Dissipative terms  can be neglected in equations (\ref{e0} and \ref{e01}).
The equation for the angular momentum conservation takes a  form
\begin{equation}
 \dot M {\partial r V_k\over \partial r}+r^2 <B_{\varphi} B_z>|_{wind}=0.
 \label{e1}
\end{equation}
Energy conservation equation is reduced to 
\begin{equation}
 {1\over 2}{\partial V_k^2\dot M\over\partial r} + 4\pi r\rho v_z E_{tot}|_{wind}=0.
 \label{e2}
\end{equation}
The luminosity of the disk is calculated by  eq. (\ref{e03}).

This system of equations together with the system of MHD equations describing the flow  of the  wind outside the disk totally defines the dynamics of the disk and the wind in the regime of ''cold`` accretion. These equations do not contain dissipative terms connected with viscosity or finite conductivity.   According to eq. (\ref{e1}),  
\begin{equation}
 <B_{\varphi} B_z>|_{wind}={\dot M V_k\over 2r^2}.
\end{equation}
This  allows us to estimate the minimum magnetic field at the base of the wind which can
 support  the regime of the ''cold`` accretion: 
\begin{equation}
 B_{min}=\sqrt{{\dot M V_k\over 2r^2}}.
 \label{bmin}
\end{equation}
It is easy to demonstrate  that this value is small compared to the field in the Shakura-Sunyaev regime of accretion. Nevertheless, the presence of the magnetic field 
at the base of the wind means that the magnetic field is present inside the disk and this field generates turbulence which makes inevitable heating and radiation from the disk. The main question for us is how strong is this radiation?

To connect the magnetic field at the base of the wind with $t_{r\varphi}$ we introduce a parameter
\begin{equation}
 \theta={4\pi t_{r\varphi}\over <B_{\varphi} B_z>|_{wind}}.
\end{equation}
Physical sense of this parameter becomes clear if to pay attention that according to the assumptions of \cite{1973A&A....24..337S} and recent numerical simulations of the magnetic field generation \cite{2016MNRAS.457..857S} 
\begin{equation}
 -t_{r\varphi} \approx 0.4 \cdot {B^2\over 4\pi},
\end{equation}
where the magnetic field $B$ is taken at the midplane of the disk.  Roughly, $\theta$ equals to the ratio of the magnetic pressure inside the disk to the magnetic pressure  at the base where the wind starts to flow. The ratio  $\theta$ depends on variables like the mass flow rate $\dot M$, mass of the central object $M$, radius,  and others. This is one of the key difference of our work from other works. We distinguish the magnetic field
inside the disk from the magnetic field at the base of the wind. In the works of the Grenoble group $\theta \sim 1$ \cite{1995A&A...295..807F,1997A&A...319..340F} and other authors \cite{2013ApJ...765..149C,2014ApJ...786....6L,2014ApJ...788...71L} $\theta \sim 1$ because   the magnetic field  vertically crosses the disk. In reality, the magnetic field inside the  disk  can essentially exceed the field at the base of the wind \cite{2016MNRAS.457..857S}.
Therefore, below we assume that  $\theta \ge 1$  but should satisfy to the condition of ''cold``accretion following from  
eq.(\ref{iq1})
\begin{equation}
 {\theta h\over r} \ll 1.
 \label{iq2}
\end{equation}
For the geometrically thin disks with $h/r \ll 1$ the ''cold`` disk accretion can take place even for $\theta > 1$.  

\section{Thermal radiation from the disk at  the ''cold`` accretion}

In the work \cite{1973A&A....24..337S},  three regimes of the disk accretion have been considered:  a) when the radiation pressure exceeds the gas pressure and the Thomson scattering dominates over the free-free absorption;  b) when the gas pressure dominates over
 the radiation pressure but the  Thomson scattering dominates the free-free absorption;  c) when the gas pressure dominates over the radiation pressure and the opacity of the matter is defined by the free-free absorption. We consider only the case when the gas pressure dominates over the radiation pressure. These are the regimes b) and c). Below we will see that when the radiation dominates, the accretion proceeds in the Shakura-Sunayev regime. 

\subsection{Scattering dominates over free-free absorption}   
Firstly, we consider the case when the Thomson scattering dominates over 
the free-free absorption. Hereafter we call this regime as Thomson regime. 
The pressure of radiation $P_{rad}$  equals to ${\ve / 3}$, where $\ve=bT^4$. The sound velocity is defined as $v_s^2= {kT/ m_p}$, where $m_p$ is the proton mass.  
According to \cite{1973A&A....24..337S} the heat conductivity of the disk is defined by the transport of radiation. Then
\begin{equation}
 \ve ={3\over 4}{Q\sigma u_0\over c},
\end{equation}
where $\sigma = 0.4 ~~\rm cm^2/g$ is the Thomson opacity and  $u_0=2\rho h$.
Then the rate of heating of the disk equals 
\begin{equation}
 Q = {3\theta\dot M V_k v_s\over 16\pi r^2}.
\end{equation}
We used here that $h=v_s/\Omega$. The solution of  this system of equations gives
\begin{equation}
T= {\sqrt{3}\over 4\sqrt{\pi}}\left({\theta^2\dot M^2 V_k\sigma\over b\alpha c r^3}\right)^{{1\over 4}}.
\end{equation}
The sound velocity equals
\begin{equation}
v_s={6^{1/4}\over 2 (2\pi)^{1/4}} {k^{1/2} V_k^{1/8}(\theta \dot M)^{1/4}\sigma^{1/8}\over m_p^{1/2}b^{1/8}\alpha^{1/8}c^{1/8}r^{3/8}} ,
\end{equation}
and the density flux of radiation from one side of the disk  is expressed as
\begin{equation}
 Q={(3\pi)^{5/4}6^{1/4}\over 32}{(\theta \dot M)^{5/4} V_k^{9/8} k^{1/2}\sigma^{1/8}\over r^{19/8} m_p^{1/2}b^{1/8} \alpha^{1/8} c^{1/8}}.
\end{equation}

Let us express  $\dot M= \dot m \dot M_{crit}$ , the radius $r$ in $r= (3 r_g)x$ and the 
mass $M$ in the solar masses $M=m M_{\sun}$. 
In these variables we obtain
\begin{equation}
 Q=0.75\cdot 10^{23}{(\theta \dot m)^{5/4}\over m^{9/8} x^{47/16}\alpha^{1/8}}, ~~\rm {erg/s/cm^2}.
\end{equation}
The integration of this expression over the disk gives the bolometric luminosity of the disk
\begin{equation}
 L_{bol}=0.84\cdot 10^{36}{(\theta\dot m)^{5/4} m^{7/8}\over \alpha^{1/8}},~~\rm {erg/s}.
\end{equation}

The kinetic luminosity of the jets equals to the total energy release at accretion. Therefore 
\begin{equation}
 L_{kin}={\dot M c^2\over 12}=1.4\cdot 10^{38} m \dot m, ~~\rm {erg/s}.
 \label{lkin}
\end{equation}
Then, the ratio of the kinetic luminosity over the bolometric luminosity equals 
\begin{equation}
{L_{kin}\over L_{bol}}= 168 {(m\alpha)^{1/8}\over \dot m^{1/4} \theta^{5/4}}.
\end{equation}
The bolometric luminosity can be expressed in conventional variables: 
\begin{equation}
 L_{bol}={4\over 5}\theta\dot MV_{k0}v_{s0},
\end{equation}
where $V_{k0}$ and $v_{s0}$ are the Keplerian and sound velocities at the inner edge of the disk.
Taking into account that the kinetic luminosity 
\begin{equation}
 L_{kin}={\dot M V_{k0}^2\over 2} , 
\end{equation}
the condition $L_{bol}/L_{kin} \ll 1$ becomes 
\begin{equation}
{8\over 5} {\theta v_{s0}\over V_{k0}}= {8\over 5} {\theta h\over r} \ll 1,
\end{equation}
which practically coincides with the condition of applicability of the ''cold`` disk accretion approximation defined by eq. (\ref{iq2}). 
Similar condition has been obtained earlier in \cite{2006A&A...447..813F}.
The condition  $L_{kin}\gg L_{bol}$  indicates that the accretion occurs in the  ''cold`` regime. 

The  temperature in  the disk 
\begin{equation}
 T=2.5\cdot 10^7 {\sqrt{\theta\dot m}\over\alpha^{1/4} x^{7/8} m^{1/4}}, ~~\rm K
\end{equation}
can be less than  the temperature in the \cite{1973A&A....24..337S} disk provided that $\theta < 100$. 

Let us calculate the ratio of the radiation pressure over the gas pressure  in the disk, 
\begin{equation}
 {P_{rad}\over P_{gas}}={3\over 32\pi} {\theta\dot M \sigma\over r c}=0.85 {\theta\dot m\over x}.
 \label{pressures}
\end{equation}

This means that all the estimates are valid at $0.85 \theta\dot m < 1$.

Other parameters are estimated in the disk as follows. Density equals
\begin{equation}
\rho = {1\over 2\sqrt{3\pi}}{\sqrt{\theta\dot M}V_k^{3/4} m_p b^{1/4}c^{1/4}\over r^{5/4}k\alpha^{3/4}\sigma^{1/4}}=  0.6{\sqrt{\theta\dot m}\over m^{3/4}x^{13/8}\alpha^{3/4}},~~ \rm g/cm^3.
\end{equation}
The aspect ratio of the disk is
\begin{equation}
 {h\over r}=3.7\cdot 10^{-3}{(\dot m\theta)^{1/4}x^{1/16}\over (\alpha m)^{1/8}}.
\end{equation}
The true optical depth $\tau^*=\sqrt{\sigma\cdot \sigma_{ff}}\cdot u_0$ of the disk is expressed as
\begin{equation}
 \tau^*=51{(\theta\dot m)^{1/8} m^{3/16}x^{5/32}\alpha^{-13/16}},
\end{equation}
where
\begin{equation}
 \sigma_{ff}=0.11\cdot T^{-7/2} n, ~~ \rm cm^2/g
\end{equation}
is the free-free opacity of the disk. 
The surface temperature of the disk $T_S$ is defined from the equation $bcT_s^4/4=Q$ and has  a form 
\begin{equation}
 T_s=6\cdot 10^6{(\theta\dot m)^{5/16}\over m^{9/32}x^{47/64}\alpha^{1/32}} ~~\rm K.
\end{equation}

\subsection{Free-free absorption dominates over scattering}

At the condition 
\begin{equation}
 4.6\cdot 10^{-3} {(\alpha m)^{1/10}x^{23/20}\over (\theta \dot m)} > 1
\end{equation}
the Thomson scattering opacity becomes small compared with  the free-free absorption. Hereafter we call this regime as free-free.
Similar calculations give the following temperature inside the disk 
\begin{equation}
 T=10^{7}{(\theta\dot m)^{6/17}\over x^{12/17}(\alpha\cdot m)^{4/17}}~\rm K.
\end{equation}
The bolometric luminosity of the disk is
\begin{equation}
 L_{bol}=0.6\cdot 10^{36}(\theta \dot m)^{20/17} m^{15/17} \alpha^{-2/17}~~\rm erg/s.
 \label{lbol}
\end{equation}
The ratio of the kinetic luminosity over the bolometric luminosity equals
\begin{equation}
 {L_{kin}\over L_{bol}}={228(\alpha m)^{2/17}\over \dot m^{3/17}\theta^{20/17}}
 \label{ratio}
\end{equation}
The full optical depth, the density of plasma and the aspect ratio of the disk  are 
\begin{equation}
 \tau=93(\theta\dot m)^{4/17}m^{3/17}x^{1/34}\alpha^{-14/17},
\end{equation}
\begin{equation}
 \rho={1.2(\theta\dot m)^{11/17}\over (\alpha m)^{13/17}\cdot x^{61/34}}~\rm g/cm^3,
\end{equation}
\begin{equation}
 {h\over r}=2.5\cdot 10^{-3}x^{5/34}(\theta\dot m)^{3/17}(\alpha m)^{-2/17}.
\end{equation}
Finally, the surface temperature   is 
\begin{equation}
 T_s=5.5\cdot 10^6{(\theta\dot m)^{5/17}\over m^{19/68}x^{97/136}\alpha^{1/34}}~~\rm K.
\end{equation}

\section{Comparison with the fundamental plane of the black holes}

The fundamental plane encapsulates the relationship between the compact radio luminosity, X-ray luminosity, and the black
 hole mass  and provides a good description of the data over a very large range of black hole
mass. There are reasons to believe that the Fundamental Plane (FP) of the black holes reproduces the actual relationship between the kinetic luminosity of jets and bolometric luminosity of the disks. In the work  \cite{2016arXiv160601399D}, the position of objects of different masses in the coordinates $L_{kin}/L_{bol}$ and $L_{bol}/L_{Edd}$ has been collected in one FP.  If this is true, the FP can be used to  extract  information about the dependence of $\theta$ on $\dot m$ and $m$.
 All  data at the FP can be approximated by a  power law function of the form 
\begin{equation}
  \log {L_{kin}\over L_{bol}}= (A-1)\log ({L_{bol}\over L_{Edd}}) +B
  \label{fp}
\end{equation}
with $A$ in the range (0.43 - 0.47) and $B$ in the range from -0.94 to -1.37. For estimates the values $A=0.457$ and $B=-1.1$ have been chosen which are close to the average. 
The ratio of $L_{bol}/L_{Edd}$ in the Thomson regime is
\begin{equation}
 {L_{bol}\over L_{Edd}}=6\cdot 10^{-3} {(\theta\dot m)^{5/4} \over (\alpha m)^{1/8}},
 \label{lbolT}
\end{equation}
while in free-free regime this ratio equals to
\begin{equation}
 {L_{bol}\over L_{Edd}}=4.4\cdot 10^{-3} {(\theta\dot m)^{20/17} \over (\alpha m)^{2/17}},
\end{equation}

Obviously,  at the constant $\theta$ the theoretical predictions are not consistent with observations. The reason is that  $\theta$ must depend on $\dot m$ and $m$.  The most natural option is to  assume that $\theta$ depends on $\dot m$ as a power  law 
\begin{equation}
\theta = D \dot m^{\gamma}.
\end{equation}
In the Thomson regime 
\begin{equation}
 X={L_{bol}\over L_{Edd}}=6\cdot 10^{-3} {\dot m^{5(\gamma+1)/4} D^{5/4}\over (\alpha m)^{1/8}},
\end{equation}
and 
\begin{equation}
 Y={L_{kin}\over L_{bol}}={168 (\alpha m)^{1/8}\over D^{5/4}\dot m^{(5\gamma+1)/4}}
\end{equation}
After simple algebraic calculations,  we obtain that 
\begin{equation}
A={4\over 5(\gamma+1)}, 
\end{equation}
For $A=0.457$  the value 
$\gamma =3/4=0.75$. Then  $D=5\cdot 10^3 (\alpha m)^{1/10}$.
Thus, in  the Thomson regime
\begin{equation}
 \theta= 5\cdot 10^3 \dot m^{3/4} (\alpha m)^{1/10}
 \label{beta}
\end{equation}
Similar calculations in the free-free regime give
\begin{equation}
 \theta= 11.4\cdot 10^3 \dot m^{0.86} (\alpha m)^{1/10} .
 \label{beta2}
\end{equation}
The power at $\dot m$ is chosen to provide  uniform dependence of $Y$ on $X$ of the form (\ref{fp}) with constant $A$ in both regimes.

The dependencies (\ref{beta}) and (\ref{beta2}) seem  physically reasonable. They show that the smaller the accretion rate, more uniform is the magnetic field across the disk and, therefore, the $\theta$ is close to 1.  

\begin{figure}
 \centering
 \includegraphics[width=0.8\textwidth]{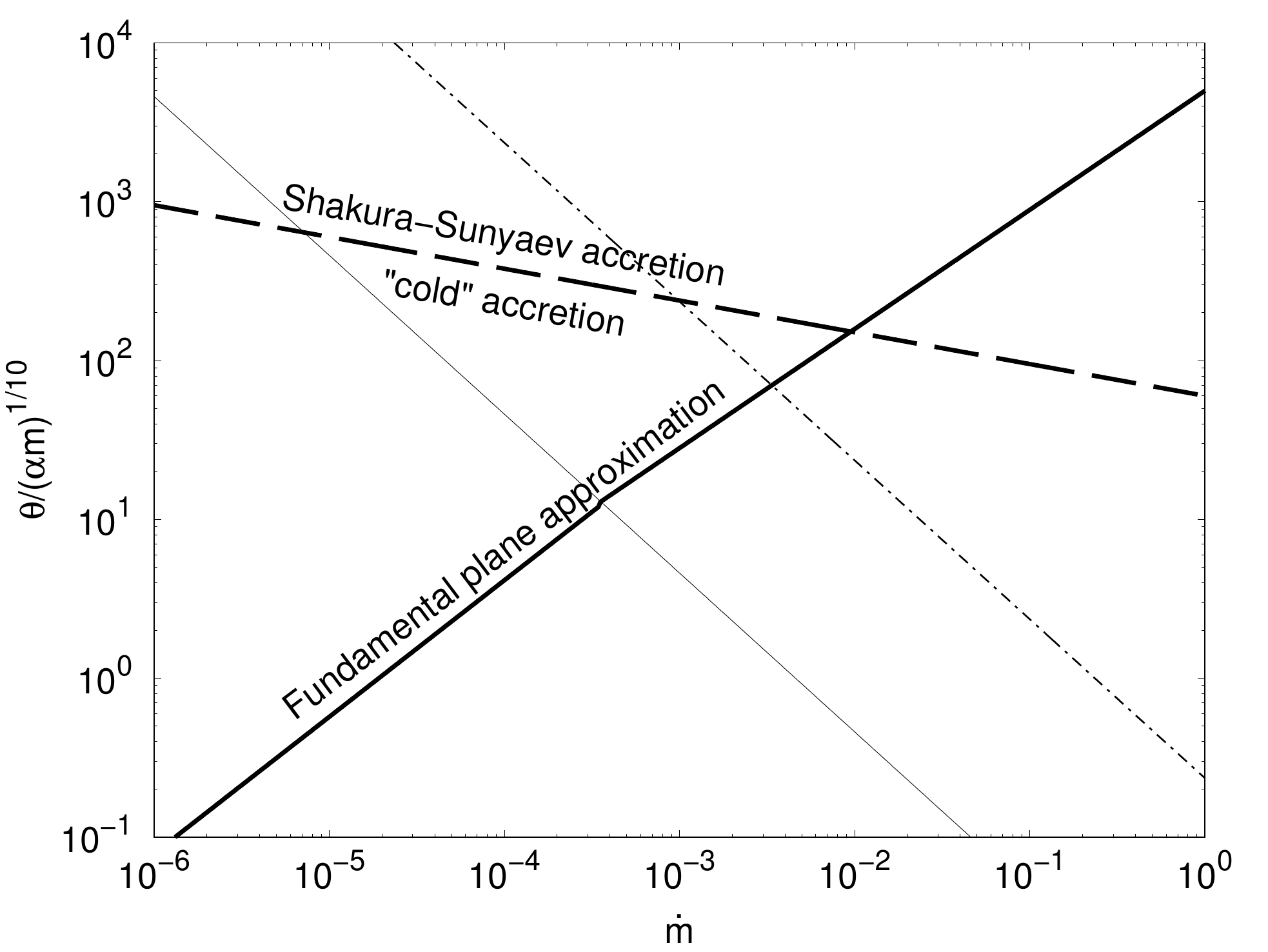}
 \caption{Dependence of $\theta/(\alpha m)^{1/10}$ on $\dot m$. 
 Shakura-Sunyaev regime of accretion takes place above thick dashed line. Below this line the regime of ''cold`` accretion takes place. 
 The Thomson scattering dominates above the thin solid line, while below this line the free-free absorption gives the major contribution into 
 the opacity of the medium. Dashed-dotted line (calculated at $m=10^8$) divides the plane in  parts where the  radiation pressure (above this line) or the gas pressure (below) dominate. }
\label{fig2}
\end{figure}

The plot of $\theta/(\alpha m)^{1/10}$ is presented in fig. \ref{fig2}.  $\theta$ corresponding to FP  agree with the assumption of the ''cold`` accretion because this curve is located well below the curve separating the regime of the cold accretion from the Shakura-Sunyaev regime. In fig. \ref{fig2} we present in dashed-dotted
the line which separates regions of domination of the gas pressure and  radiation pressure  defined by equation (\ref{pressures}). Thin solid  line separates the  Thomson regime from the free-free regime.

\section{Comparison with the specific sources}
It is interesting to apply the estimated dependencies to the specific sources. Below we consider  M87 and the SMBH in galactic center,  Sgr A*.  We will see later that 
both sources are in the free-free regime. Therefore eq. (\ref{beta2}) has been used at the estimations. 
For definiteness  we accept $\alpha = 0.1$.
\subsection{M 87}
For this object $\dot m$ and $m$ can be easily estimated.  The kinetic luminosity of this object is $L_{kin}=10^{44}~\rm erg/s$ which we assume is equal to the total rate of the gravitational energy released at the accretion. 
Mass of the central black hole $m=3.5\cdot 10^9$ \cite{2013ApJ...770...86W}. With the Eddington luminosity equal  
to $L_{Edd}=1.4\cdot 10^{38}m ~\rm erg/s$,  we find  $\dot m = L_{kin}/L_{Edd}=2\cdot 10^{-4}$. From eq. (\ref{ratio}) we obtain that
\begin{equation}
 {L_{kin}\over L_{bol}}=10^4 \theta^{-20/17}.
\end{equation}
Eq. (\ref{beta2}) gives  $\theta=54$. Then  ${L_{kin}/L_{bol}} \approx 95$ and 
$L_{bol}$ from eq. (\ref{lbol}) equal to $10^{42} ~\rm erg/s$ in accordance with observations. Optical depth of the disk exceeds $\tau > 10^4$. 

\subsection{Sgr A*}
The kinetic luminosity of the outflow  from the disk around SMBH  in Sgr A* is not known.  The flux of TeV gamma-rays from the Galactic Center can be explained  by very high energy accelerated protons   with a luminosity close to $10^{38}~\rm erg/s$. The kinetic luminosity of the wind has to be higher.  We consider a reversed problem. Assuming that the dependence (\ref{beta2}) is valid for the disk in Sgr A*  we can predict what is the kinetic luminosity of wind from the disk in the Galactic Center.  Let us to estimate $\dot m$ from the bolometric luminosity of the disk (see eq. (\ref{lbol})). In this case
\begin{equation}
 \dot m=\left(({L_{bol}\over 3.6\cdot 10^{40}~({\rm erg/s})})^{17/20} {1\over m}\right)^{1\over 1.86}=8\cdot 10^{-6},
\end{equation}
$\theta=1.7$ and from eq. (\ref{lkin}) we obtain that
\begin{equation}
 L_{kin}=4.4\cdot 10^{39},~\rm erg/s,
\end{equation}
at $L_{bol} = 10^{36}~\rm erg/s$. 
The kinetic luminosity of the wind from the Galactic accretion disk $4.4\cdot 10^3$  times exceeds the bolometric luminosity of the disk. 
Remarkably, this power is sufficient to explain the flux  of the PeV protons in the Galactic Center.

\section{Discussion}

The main result of this work is the conclusion that the main source of energy in AGN is the energy released at the accretion even in the case of 
high ratio of the kinetic-to-bolometric luminosity. Observation of such objects can point out that the accretion occurs in the ''cold``regime. 
Outflow in the form of a wind  from the accretion disk  carry out the largest fraction of the angular momentum of the accreted material.
This results into suppression of radiation from the disk. Accretion onto SMBH appear very efficient process which transforms practically all the gravitational 
energy into the kinetic energy of jets. This strongly contrasts with also radiationally inefficient  ADAF   models \cite{1997ApJ...489..865E}. In the last case 
AGN appears very inefficient machine which  transforms almost all the gravitational energy into mass and rotational energy of the SMBH. 
This model needs  an additional source of energy which is believed can be the rotational energy of SMBH. 

The accretion in the ''cold`` regime occurs
even when the magnetic field inside the accretion disk essentially exceeds the magnetic field  at the base of the wind. This is explained by the geometrical reason. The angular momentum transport due to viscosity is proportional to the magnetic pressure in the disk times the thickness of the disk $h$ while the flux of the angular momentum from the disk is proportional to the magnetic pressure at the base of the wind times  the radius.  The ratio of the viscous losses to the losses due to wind is $\sim \theta\cdot({h\over r})$, where $\theta$ roughly equals to the ratio of the magnetic pressures inside and at the surface of the disk. Therefore, the Shakura-Sunyaev regime of accretion \cite{1973A&A....24..337S}  is realized  when $\theta \gg ({r\over h})$.  The  assumption that $\theta$ is constant contradicts to the observations  if to assume that  FP correctly reproduces the ratio of the kinetic-to bolometric luminosities. $\theta$ increases on 2 orders of magnitude  with $\dot m$.
  This value of $\theta$ agrees with the  results of modeling of MRI dynamo \cite{2016MNRAS.457..857S}. 
 This  estimate  should be considered as only a rough approximation. Nevertheless, it allows us to make certain conclusions about realization of the regime of ''cold`` disk accretion. It appears that at small accretion rates $\dot m < 10^{-2}$, the estimated value of $\theta$ is located in the region well below the line were the Shakura-Sunyaev model is valid. The magnetic pressure inside the disk appears  less than the magnetic pressure estimated in the model   \cite{1973A&A....24..337S}. Therefore, it is quite reasonable to assume that  at relatively low rates of accretion, $\dot m <  10^{-2}$, 
the accretion occurs predominantly in the regime of the ''cold`` accretion. At higher values of $\dot m > 0.1$ the accretion  occurs in the regime of Shakura \& Sunyaev. 
The transition between the regimes takes place at the value of $\dot m$ between $0.01$ and $0.1$ which well agree with location of the transition 
from very bright to very dim disks around SMBH with powerful outflow deduced in \cite{2005MNRAS.363L..91C}.
Remarkably, the rough estimate of the dependence of $\theta$ on $\dot m$ gives good agreements  with observations of  two   SMBHs, M87 and  Sgr A*.  Hardly, such an agreement could be accidental. 

In this paper, we have considered only the case when the gas pressure dominates over the radiation pressure. But according to fig. \ref{fig2},   the radiation pressure dominates basically in  the Shakura-Sunyaev accretion regime.  The ''cold`` accretion at the domination of the radiation pressure  apparently may be realized in rather narrow range of parameters. This case is planned to be considered separately.  

\section{Acknowledgments}

\begin{acknowledgments}
The  work was supported by Russian science  foundation, project N 16-12-10443. Author is grateful to Felix Aharonian for stimulation of this work and 
comprehensive discussions. Grateful also follow to N.I.Shakura for the discussion 
of realization of different accretion regimes,  and to M.R.Gilfanov  who called  my attention to the 
information that  can be extracted from Fundamental Plane of black holes. 
\end{acknowledgments}

%

\end{document}